%% file: tmsl.tex
\documentclass[aps, prd, amsmath, amssymb, reprint, groupedaddress]{revtex4-1}

\usepackage{graphicx}
\usepackage{dcolumn}
\usepackage{bm}
\usepackage{xcolor}
\usepackage{physics}
\usepackage{textcomp}
\usepackage{siunitx}
\usepackage{amsmath} 
\usepackage{verbatim} 
\usepackage{color}    
\usepackage{subfigure} 
\usepackage{hyperref}
\usepackage{float}
\usepackage[normalem]{ulem}

\newcommand{\gl}[1]{\textbf{\textcolor{red}{#1}}}

\begin{document}

\preprint{AIP/123-QED}

\title[]{Noise reduction in qubit readout with a two-mode squeezed interferometer}

\author{G. Liu$^\dag$, X. Cao$^\dag$, T.-C. Chien, C. Zhou, P. Lu, and M. Hatridge}
\email{gangqiang.liu@yale.edu; hatridge@pitt.edu}

\thanks{$^\dag$ these authors contributed equally}

\affiliation{Department of Physics and Astronomy, University of Pittsburgh, Pittsburgh, Pennsylvania 15260, USA}

\date{\today}

\begin{abstract}
\textbf{Abstract}
Fault-tolerant quantum information processing with flawed qubits and gates requires highly efficient, quantum non-demolition (QND) qubit readout.  In superconducting circuits, qubit readout using coherent light with fidelity above 99\% has been achieved by using quantum-limited parametric amplifiers such as the Josephson Parametric Converter (JPC). However, further improvement of such measurement is fundamentally limited by the vacuum fluctuations of the coherent light used for readout. In this work we measure a transmon qubit/cavity system with an unbalanced two-mode squeezed light interferometer formed from two JPCs. The first amplifier generates two-mode squeezed vacuum at its output, which is coherently  recombined by the second amplifier after one branch is shifted and displaced by the transmon's state after it interacts with the qubit/cavity system on one arm of the interferometer.  We have observed a 31\% improvement in power Signal-to-Noise Ratio (SNR) of projective readout compared to that of coherent light readout in the same system.  To investigate the quantum properties of the two-mode squeezed light in the system, we also studied weak measurement and found, surprisingly, that tuning the interferometer to be as unprojective as possible was associated with an increase in the quantum efficiency of our readout relative to the optimum setting for projective measurement. These enhancements may enable remote entanglement with lower efficiency components in a system with qubits in both arms of the interferometer.

\end{abstract}


\maketitle

\section {Introduction}

For quantum computing an essential requirement for all tasks including qubit state preparation, gate calibration and error detection for quantum error correction (QEC), is the ability to perform fast single-shot qubit measurement with high fidelity. Furthermore, for error correction and state preparation, the measurement should also be quantum non-demolition (QND) -- the qubit state should not suffer additional back-action or randomization during the measurement process \cite{Caves1980}. In superconducting circuits, these requirements are fulfilled by the dispersive readout method. In this method, a pulse of coherent microwave light is sent into a cavity that is dispersively coupled to a qubit inside it \cite{Blais2004, Clerk2010}. The pulse experiences a qubit-state-dependent phase and/or amplitude shift which entangles the state of the pulse with the qubit. The state of the qubit, is determined by measuring the pulse at room temperature after the signal is amplified by a chain of low noise amplifiers.

Fast high fidelity single-shot dispersive readout has been achieved in superconducting circuits thanks to the use of Josephson junction based parametric amplifiers (JPAs) as the first amplifier in the amplification chain \cite{Yamamoto2008, Castellanos-Beltran2008, Bergeal2010a, Frattini2018}. These amplifiers provide high gain while adding only the minimum amount of noise to the signal required by quantum mechanics. For a given pulse amplitude, the signal-to-noise-ratio (SNR) of the readout is thus fundamentally limited by the quantum fluctuation of the measurement light and the added noise of the JPA, which is half a photon for phase-preserving amplification \cite{Caves1982}. Consequently, large SNR and high fidelity can be achieved in single-shot qubit readout with pulses containing just a few microwave photons. In the meantime, the QND requirement is satisfied as the readout photon number is much smaller than the critical photon number at which the dispersive coupling breaks down \cite{Clerk2010}.

Although the state-of-art of dispersive readout is sufficient for small quantum circuits, the implementation of large quantum circuits (such as a error corrected machine) requires high fidelity readout as the probability of measuring all qubits correctly in a circuit decreases exponentially with the number of qubits. It might seem straight forward to improve readout SNR by increasing the readout strength, however, this can lead to other unwanted effects which degrade the readout quality and qubits' coherence. One well-known yet poorly understood effect is the reduction of qubit $T_1$ with increasing measurement strength, which leads to reduction in measurement fidelity and QNDness \cite{Boissonneault2009, Verney2019, Lescanne2019, Petrescu2020, Malekakhlagh2020, Hanai2021, Shillito2022}. Such effect has been observed in readout with as few as 2.5 photons \cite{Vool2014, Walter2017, Minev2019}. 

In this work, we have focused on circumventing the quantum mechanical limits on noise in qubit measurement.  One method to achieve this is to use squeezed light to reduce noise inside the signal to achieve higher SNR. Single-mode squeezed light has long been used in quantum metrology, in particular in interferometry of which the most famous example is the gravitational wave detector, to achieve measurement precision beyond the standard quantum limit set by the quantum fluctuations in coherent states of light \cite{Caves1980, Caves1981, Giovannetti2004, Goda2008, Abadie2011, Aasi2013}. 
It is therefore natural to expect that dispersive qubit readout could also be improved by using single-mode squeezed light. However, dispersive coupling will induce a qubit-state-dependent rotation of the squeezing axis, therefore mixing amplified noise of the anti-squeezed quadrature into the squeezed quadrature of the single-mode squeezed light \cite{Barzanjeh2014}. Furthermore, optimal dispersive readout typically requires large rotation. Therefore, only modest SNR improvements have been achieved in qubit readout with single-mode squeezed light instead \cite{Eddins2018}.

\begin{figure*}[t]
	\includegraphics[scale = 1.0]{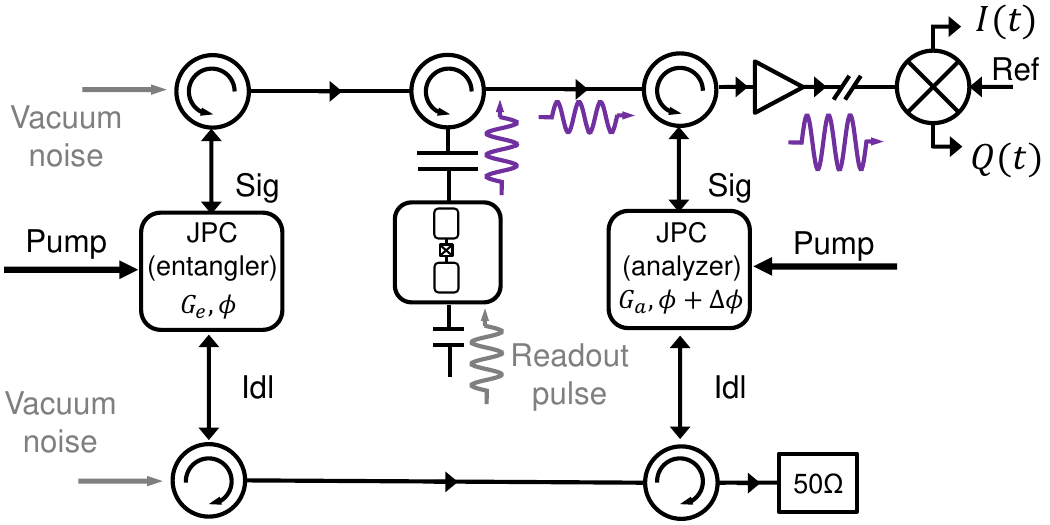}
		\caption{ \textbf{Schematic of the experiment setup.} Two nominally identical JPCs, entangler and analyzer, are connected by their signal (Sig) and idler (Idl) ports respectively with circulators and superconducting cables to form an interferometer. The unused ports on the circulators are terminated in cold 50~$\Omega$ loads which provide quantum vacuum noise to the entangler, and a dump for the unused signal on the lower arm of the interferometer. The upper arm of the interferometer is interrupted by a 3D transmon qubit-cavity system. The qubit measurement pulse is sent into the cavity via its weakly coupled port without going through both arms of the interferometer which enables the switching between the two-mode squeezed light readout and coherent light readout {\it in situ} by turning on and off the pump of the entangler JPC. The signal is subsequently further amplified by a cryogenic HEMT  and demodulated and recorded at room temperature.}
\label{fig:Experiment_setup}
\end{figure*}
Here we demonstrate inteferometric qubit readout based on two-mode squeezed (TMS) light input states and phase-preserving amplification \cite{Barzanjeh2014}. This measurement scheme is based on a Mach-Zehnder interferometer with active non-linear beam splitters which is commonly known as a SU(1,1) interferometer \cite{Yurke1986}. Two-mode squeezed light naturally forms a pair of Gaussian, entangled states which travel along two paths, and when recombined in a phase-preserving amplifier, which in the quantum limit naturally has two input and output ports \cite{Caves1982, Bergeal2010}, can achieve superior readout SNR (and hence fidelity) by lowering the output noise level through the destructive interference between the two modes. In contrast to qubit readout with single-mode squeezed light, this readout scheme is more broadly compatible with the dispersive qubit-cavity coupling and allows SNR improvement over a larger range of coupling strengths.  

In the experiment, we take advantage of the fact that two-mode squeezing and phase-preserving amplification are the same process, using two nominally identical, single-ended Josephson Parametric Converters (JPCs) to generate and process two-mode squeezed light in the interferometer \cite{Bergeal2010, Bergeal2010a, Flurin2012}. The qubit-cavity system is embedded in one arm of the interferometer, where it imbalances the system by imparting a qubit-state dependent phase shift on light passing through that arm.  To allow fair comparison with standard coherent pulse readout, we perform readout by a combination of TMS vacuum and a cavity displacement drive which enters the system through the cavity's weakly coupled port, and so is subjected to no interference effects.  Thus, our overall readout scheme has fixed measurement strength and gain (20 dB), and only the noise is affected. 

We have reduced output noise power below the standard quantum limit by $40\%$ through the destructive interference between the correlated TMS noise entering the JPC \cite{Caves1982, Caves2012}, resulting in a $31\%$ improvement in power SNR comparing to coherent light readout. More, we explored the back-action and SNR of weak measurements \cite{Hatridge2013} in our TMS readout scheme. We found, to our surprise, that the observed quantum efficiency of measurement (that is the fraction of the qubit back-action we can extract from the measurement record) \cite{Carmichael1993,Wiseman2009,Korotkov2016,Hatridge2013,Bultink2018} varied inversely with noise in the interfometer, so that 'noisier' measurements were more faithful.  While we lack a detailed theory to explain this effect, it has important implications for future work on multi-qubit quantum measurements which play a central role in tasks such as error correction and entanglement generation.

\section {Experiment Setup}

In our experiment the two-mode squeezed light is generated, analyzed and amplified with JPCs (Fig.~\ref{fig:Experiment_setup}). A JPC is a non-degenerate phase-preserving amplifier based on the three-wave mixing process in a ring of four nominally identical Josephson junctions, known as a Josephson Ring Modulator (JRM). The JPC's Hamiltonian can be written as \cite{Bergeal2010, Bergeal2010a}:
\begin{equation}
\frac{H_{JPC}}{\hbar} = \omega_a a^{\dag} a + \omega_b b^{\dag} b + \omega_c c^{\dag} c + g_3 (a^{\dag} b^{\dag} c + a b c^{\dag}),
\label{eq:H_JPC}
\end{equation}
where $a$, $b$, and $c$ are the annihilation operators of the three modes of the JPC which are refereed to as `signal', `idler', and `common' mode respectively,  and $g_3$ is the three-wave coupling strength. Phase-preserving amplification is achieved by applying a strong microwave drive to the common mode at frequency $\omega_p \simeq \omega_a + \omega_b$. When $\omega_p$ is far detuned from $\omega_c$, the pump is `stiff' in the sense that $c$ can be replaced by its average value $\expval{c}e^{i\phi_p}$, where $\expval{c}$ and $\phi_p$ are the pump amplitude and phase, respectively.  This reduces the three-wave mixing term to an effective two-wave coupling term between the signal and idler mode:  {$g_3\expval{c}(e^{i\phi_p}a^{\dag} b^{\dag} + e^{-i\phi_p} a b)$}. As an amplifier, this device amplifies signals in reflection with amplitude gain of $\sqrt{G}$, and in transmission with frequency conversion and amplitude gain of $e^{i\phi_p}\sqrt{G-1}$. 

The time evolution operator arising from phase-preserving gain is the so-called two-mode squeezing operator at a certain time \cite{Caves1985}: $S = \text{exp}(r e^{-i\phi_p} a b + r e^{i\phi_p} a^{\dag} b^{\dag})$, where $re^{i\phi_p}$ is the complex squeezing parameter with $r = \cosh^{-1}{\sqrt{G}}$ \cite{Gerry1995, Loudon1987}.
Thus, the JPC, or any other non-degenerate phase-preserving amplifier, can also serve as a generator of two-mode squeezed light with both spectral and spacial non-degeneracy \cite{Bergeal2012, Flurin2012}. 

In our experiment, we form an active interferometer with two nominally identical JPCs, the `entangler' and `analyzer', by directionally connecting their signal and idler ports \cite{Flurin2012} (Fig.~\ref{fig:Experiment_setup}). 
Uncorrelated vacuum noise enters the interferometer via the inputs of the entangler JPC, which transforms them into highly correlated, two-mode squeezed vacuum traveling along the two arms of the interferometer. These two paths recombine and interfere with each other in the analyzer JPC, generating outputs controlled by the gains and relative pump phase of the two JPCs. If both JPCs' pumps are applied with zero relative pump phase the signal is simplify amplified twice, while a $180^{\circ}$  relative pump phase will cause the analyzer JPC to de-amplify the output of the entangler JPC. In the absence of loss and added noise, the output of the interferometer will return to uncorrelated vacuum if the gains of the two JPCs are matched and their pump phases are different by $180^{\circ}$.  

We use the interferometer to read out a qubit by interrupting the upper arm with a microwave cavity, which in turn is dispersively coupled to a transmon qubit, as shown in Fig.~\ref{fig:Experiment_setup}.  To achieve a superior SNR compared to qubit readout with coherent state input and phase-preserving amplification (CS + PP), the dispersive phase shift due to interaction with the qubit-cavity system must be either close to 0 or $\pi$, which correspond to the qubit-cavity dispersive shift $\chi$ being much smaller or larger than the cavity linewidth $\kappa$ \cite{Barzanjeh2014}. In our experiment, we design the quit-cavity system to have $\chi/\kappa = 0.19$, which satisfies this requirement while also being favorable for fast readout (see Section \ref{sec:Exp_params} for all the parameters of the experiment).

As we will discuss later, the noise produced by the interferometer when operated with no cavity displacement drive can itself carry qubit information  for many combinations of entangler and analyzer gain and relative phase. However, high SNR readout requires a coherent drive to be applied to the cavity. The original theory proposal called for displacing the input to the upper arm of the entangler \cite{Barzanjeh2014}, however in this scenario the signal is both amplified and partially transmitted down both arms of the interferometer, so that there is interference in both the output signal and noise.  This greatly complicates fair comparison with CS + PP readout, and so in our experiment we drive our readout through a second, weakly coupled port in the microwave cavity.
The readout signal thus does not experience any form of interference, and experiences the same gain (from the analyzer only) in both our TMS and CS + PP readout.  In this case, we squeeze only on the noise, with a degradation/enhancement of SNR corresponding to a larger/smaller output noise, respectively.

\section {Results}
\subsection{Generation and characterization of two-mode squeezed light}\label{sec:TMS_noise_exp}

As a first step towards qubit readout with two-mode squeezed light, we demonstrate the generation and detection of such a non-classical state \cite{Flurin2012}. As in previous demonstrations, the two-mode squeezed light is generated by displacing the two-mode squeezed vacuum produced by the entangler with a coherent drive applied to the weak port of qubit-cavity on the upper arm of the interferometer. The detection of the two-mode squeezed light is done by recording the output voltage from the signal port of the analyzer at room temperature as a function of the entangler pump conditions. As discussed earlier, only the noise of the output signal is expected to vary with the entangler pump, while the average amplitude should remain constant.

\begin{figure}[htbp]
	\includegraphics[scale = 1.0]{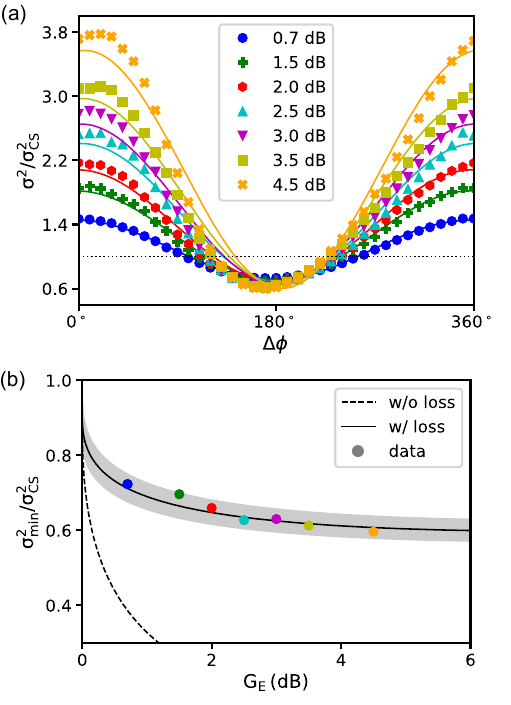}
		\caption{\textbf{Noise reduction in two-mode squeezed light.} (a) Normalized output noise power of two-mode squeezed light as a function of relative pump phase $\Delta \phi$ for different entangler gain. The black dash line represent the output noise level of the coherent vacuum (CS) input as a reference while the other colors each represents a different entangler gain. Colored lines are fits to the data by our mode (Eq.~\ref{eq:TMS_noise_norm}) with analyzer gain as the fitting parameter. (b) Calculated optimal noise reduction without (dashed line) and with photon loss (solid line) for analyzer gain of 19.4~dB. The shaded region represent results for analyzer gain of $19.4 \pm 0.3$~dB. The colored dots represent optimal noise reduction extracted from (a).}
\label{fig:TMS_noise}
\end{figure}

Fig.~\ref{fig:TMS_noise}(a) shows the variance ($\sigma^2$) of the output voltage signal, which is proportional to the noise power, normalized to that of unsqueezed inputs (coherent state readout) ($\sigma^2_{CS}$) for entangler gain $G_E$ between 0.7~dB and 4.5~dB and relative pump phase between $0^\circ$ and $360^\circ$. The analyzer is operated with the pump condition that generates 20~dB gain when the entangler is off. This analyzer gain is chosen such that the noise power of its output signal is much stronger than those of the following amplifiers. The qubit is always prepared in the ground state ($\ket{g}$), therefore it only contributes a constant phase shift to the total relative phase between the two arms of the interferometer. 

As expected, the noise power of the output signal oscillates with relative pump phase and goes below that of the amplified coherent state noise power (indicated by the dashed line) when the two pumps are out of phase. This oscillatory behavior, in particular the reduction of it by as much as $40\%$ or 2.2~dB compare to that of the coherent state, is a direct manifestation of the entanglement between photons in the two arms of the interferometer. It is worth mentioning that the average amplitude of the output signal, as shown in Fig.~\ref{fig:fitted_Ga_all}(a), also oscillates with the relative pump phase by  $\pm 5\% $ rather than remaining constant. However, this phenomenon is independent of two-mode squeezing. It is due to analyzer gain variation caused by interference between its pump tone and the entangler pump leakage. This effect is analyzed in detail in SI Section~\ref{sec:pump_leakage}. 
 
We fit these data with a model based on Ref.~\cite{Flurin2012}, which takes into account photon loss and noise added by later amplifiers, with analyzer gain being the only fitting parameter. These fits are shown in Fig.~\ref{fig:TMS_noise}(a) as curves with the same color as their respective data set. For entangler gain below 3.5~dB, this model accurately captures the noise performance. At higher entangler gain, gain compression and other higher order effects in analyzer becomes important causing deviation between data and this model (see Fig.~\ref{fig:fitted_Ga_all}). Measured optimal noise reduction ($\sigma^2_{min} / \sigma^2_{CS}$) as a function of entangler gain is shown in Fig.~\ref{fig:TMS_noise}(b) (colored dots). It agrees well with the model prediction with analyzer gain of $19.4 \pm 0.3$~dB (filled region). Without photon loss and noise from amplifiers after the analyzer, an optimal noise reduction by $90\%$ could be achieved with entangler gain of 4~dB and analyzer gain of 19.4~dB (Fig.~\ref{fig:TMS_noise}(b), dash line). Although the achievable noise reduction is limited to $40\%$, it can translate into sizeable SNR improvement and interesting measurement back-action in qubit readout as we will show in the rest of the paper.

\subsection{Qubit readout with two-mode squeezed vacuum}
\begin{figure}
	\includegraphics[scale = 1.0]{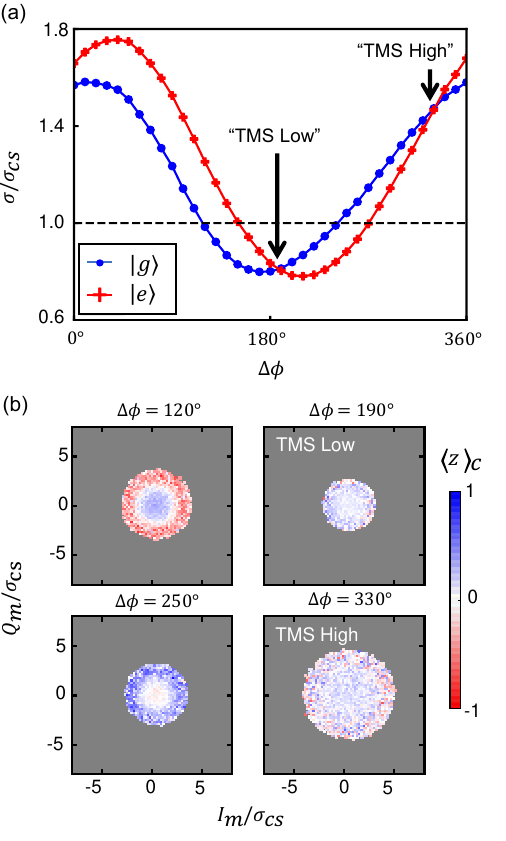}
		\caption{ \textbf{Qubit readout with two-mode squeezed vacuum.} (a) Normalized standard deviation of output noise voltage for two-mode squeezed vacuum input on the analyzer as a function of relative pump phase and qubit state. The black dotted line shows the standard deviation when the input of the analyzer is the coherent (unsqueezed) vacuum. Blue and red curves represent results for two-mode squeezed vacuum noise recorded when the qubit is in $\ket{g}$ and $\ket{e}$ state respectively. (b) Conditional tomography data for the $z$ component of the qubit Bloch vector after we record the output noise for 660~ns at different relative pump phases. The qubit is prepared in the state of $(\ket{g} + i\ket{e})/\sqrt{2}$. The entangler gain is $G_E = 2.0$~dB and the analyzer gain is $G_A = 20$~dB.}
\label{fig:Bull_eye}
\end{figure}

In order to utilize the two-mode squeezed light for qubit readout, we first study the effect of the dispersive qubit-cavity phase shift on the output noise of the interferometer. We repeat the measurements in Fig.~\ref{fig:TMS_noise} with $G_E=2.0$~dB with the qubit now being prepared either in the ground ($\ket{g}$) or excited state ($\ket{e}$), and show the results in Fig.~\ref{fig:Bull_eye}(a). The oscillation of output noise with relative pump phase for qubit in ground and excited states are very similar, with a relative phase shift of $\ang{40}$ which is due to the qubit-state dependent dispersive phase shift on photons traveling on the upper arm of the interferometer. For a given relative pump phase, this extra phase shift creates a qubit-state dependent output noise power. This means that, except for the two relative phases ($\Delta \phi = \ang{190}$ and $\ang{330}$) where the output noise is identical for both qubit states, the output noise of the interferometer can measure/dephase the qubit state without explicitly displacing the cavity field which is in sharp contrast with disperive readout with CS + PP.

To demonstrate this dispersive qubit readout with only two-mode squeezed vacuum in the interferometer, we modify our measurement protocol by changing the initial state of the qubit to $(\ket{g} + i\ket{e})/\sqrt{2}$, and adding a strong  measurement after recording the noise output voltage to determine the state of qubit after its interaction with the two-mode squeezed vacuum. As in References~\cite{Hatridge2013,Roch2014}, we construct a histogram where each pixel contains the average of all final measurement results of the $z$-component of the qubit state Bloch vector conditioned on receiving a particular voltage ($I_m$,$Q_m$) in the measurement of the output noise.  

In Fig.~\ref{fig:Bull_eye}(b), we show the conditional $z$-axis tomography results at four different relative pump phases: $\Delta \phi = \ang{120}$ and $\ang{250}$, at which the difference between the output noise power for qubit in ground and excited state is the largest, and $\Delta \phi = \ang{190}$ and $\ang{330}$ at which the output noise power for qubit in ground and excited state are the same. At $\Delta \phi = \ang{120}$, we clearly see a `bullseye' pattern, with the qubit found to be in $\ket{e}$ if the recorded noise voltage is large, and in $\ket{g}$ if the recorded noise voltage is small.  A similar result is also seen at $\Delta \phi = \ang{250}$, with the correspondence between the noise voltage and qubit states reversed.  

These results show that two-mode squeezed vacuum in our inteferometer, unlike unsqueezed vacuum, can entangle with the qubit state. An observer with a power meter could perform a readout of the qubit simply by measuring how much noise the circuit emits. This on average works poorly, as the two gaussian distributions for the qubit states share a center, and so common outcomes near the origin poorly distinguish the qubit states (hence the pale central colors in the bullseyes).  However, for outcomes far from the center the contrast between the two distributions grows exponentially; for these outcomes the qubit state is certain. The overall scheme closely resembles optical-frequency single-photon detection, in which a 'click' outcome gives a definite result, while no-click is ambiguous. This also implies that powering the entangler and analyzer will generate continuous qubit dephasing at these bias points.  In contrast, at $\Delta \phi = \ang{190}$ and $\ang{330}$ where the output noise levels are the same for different qubit states, similar to unsqueezed vacuum, no information about the qubit can be inferred from the system's noise output. 
 
\subsection{Qubit readout with two-mode squeezed light}
 We next show that SNR of such measurements can be improved by using a fixed measurement tone and replacing the un-squeezed noise in the interferometer with two-mode squeezed fluctuations. The power SNR in our experiment is defined as:
\begin{equation}
\text{SNR} = \frac{(I_c^g - I_c^e)^2 + (Q_c^g - Q_c^e)^2}{\sigma_g^2 + \sigma_e^2}
\label{eq:SNR}
\end{equation}
where $(I_c^{g(e)}, Q_c^{g(e)})$ is the center of the measurement result distribution when the qubit is in the  $\ket{g}$ ($\ket{e}$) states and $\sigma_{g(e)}$ are the corresponding standard deviations. 

To determine the SNR, we prepare the qubit in the ground and excited state separately, and then perform readout by sending a coherent probe signal through the cavity from its weak port. Given that the coupling strength of the strong port is much greater than that of the weak port $(Q_{weak} \gg Q_{strong})$, quantum fluctuation of the cavity field will be set by either two-mode squeezed vacuum (entangler on) or vacuum (entangler off) entering from the strong port. With this setup, it is straight forward to obtain fair comparison with standard CS + PP readout as there is ideally no interference in the average amplitude of the readout signals, and so any change in SNR will be solely due to changing the quantum noise in the interferometer. 

\begin{figure}
	\includegraphics[scale = 1.0]{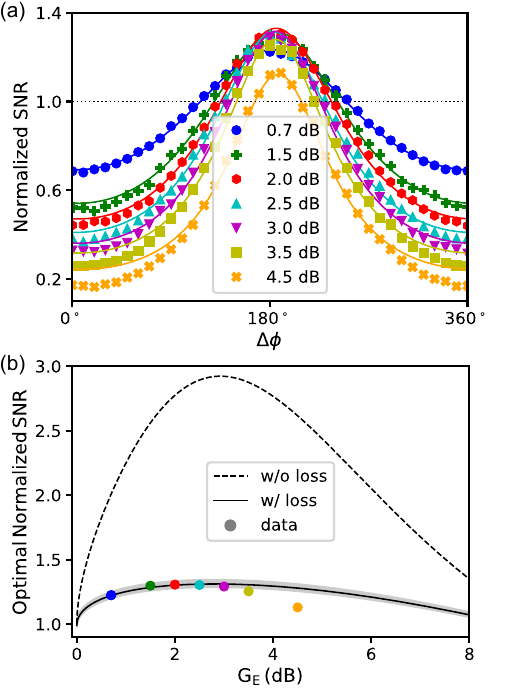}
		\caption{\textbf{SNR of qubit readout with two-mode squeezed light.} (a) Normalized SNR of qubit readout with two-mode squeezed light  for different entangler gain as a function of relative pump phase. The colored symbols represent experiment data. The colored lines represent fitting to the data with our model (Eq.~\ref{eq:TMS_SNR_norm}). (b) Calculated optimal normalized SNR without (dashed line) and with (solid line) photon loss for analyzer gain of 19.8~dB. The shaded region represents calculation results for analyzer gain between 19.5~dB (lower boundary) and 20.2~dB (upper boundary). Colored dots are optimal normalized SNR extracted from data set shown in (a) with the same color.} 
\label{fig:SNR}
\end{figure}

\begin{figure*}
	\includegraphics[scale = 1.0]{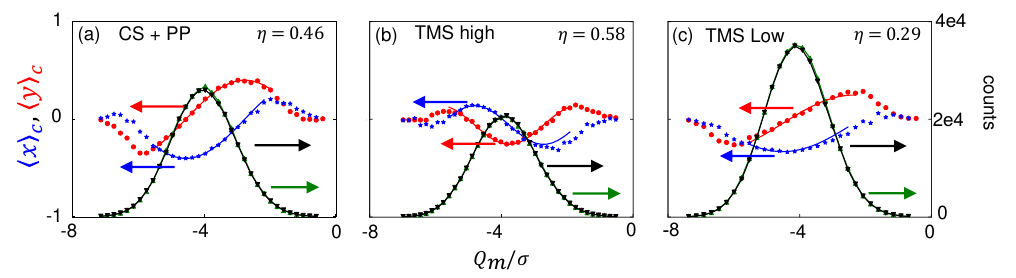}
		\caption{\textbf{Quantum efficiency obtained by analyzing the back-action of weak measurements} Tomography data for x and y (blue stars and red dots) components of the qubit Bloch vector after we apply a weak measurement (see supplementary for details) are plotted against the left axis. The counts observed for each x and y tomography data points are shown with black down pointing triangle and green up pointing triangle, plotted against the right axis. The data is recorded for coherent light as well as two-mode squeezed light at relative pump phase point such that the ground and excited state qubit has the same noise level. Data is then fitted to the theory model for coherent light \cite{Hatridge2013}. The measurement strength is adjusted such that all the measurements have the same SNR. The efficiency value obtained for each point is shown as well. The $\eta$ is the quantum efficiency extracted from the fitting model. A surprisingly low quantum efficiency is observed when the system has less output noise level.}
\label{fig:Quantum efficiency}
\end{figure*}

In this experiment, the cavity drive strength is the same as that used Fig.~\ref{fig:TMS_noise} which populates the cavity with 2.1 photons in steady state and gives a SNR of 0.9 in CS + PP readout. With this cavity drive strength, entangler gain as large as 3.5~dB can be used without causing significant analyzer gain compression (see Fig.~\ref{fig:TMS pump leakage}(b)). This allows us to perform qubit readout with the maximum amount of noise suppression achievable in our system (Fig.~\ref{fig:TMS_noise}). It also allows us to characterize the measurement efficiency of these readout schemes with the weak measurement protocol under same conditions which will be discussed in the next section. 

Fig.~\ref{fig:SNR}(a) shows the SNR of two-mode squeezed light readout as a function of relative pump phase ($\Delta \phi$) for different entangler gains. The data is normalized to the average SNR of CS + PP readout ($G_E=0$~dB) with the same cavity drive strength. Improvement in SNR, as large as $31\%$, is observed for relative pump phase in the range of $\ang{140}$ to $\ang{240}$. Outside this range, readout with two-mode squeezed light has a lower SNR than with coherent light due to one or both of the two qubit states having a substantially higher noise than coherent light readout. These data are accurately reproduced (color curves) by a fit to our model with analyzer gain as a fit parameter (Section~\ref{sec:TMS_thry}), with analyzer gain between 19.5~dB and 20.2~dB for as expected $G_E \leq 3.5$~dB, and 17.9~dB for $G_E = 4.5$~dB which accounts for analyzer gain compression under this entangler gain.

Optimal SNRs extracted from Fig.~\ref{fig:SNR}(a) are shown in Fig.~\ref{fig:SNR}(b) with respect to entangler gain. The dependence of optimal SNR on entangler gain is captured by our mode with analyzer gain of $19.8 \pm 0.3$~dB (shaded area) except for $G_E=4.5$~dB at which analyzer gain is reduced to 17.9~dB due to gain compression. As shown by both data and theory, the optimal SNR improvement first increases to a maximum value of $31\%$ for entangler gain between 2.0~dB and 3.0~dB, then decreases at higher entangler gain. This behavior results from the competition between better noise reduction at optimal pump phase -- $170^\circ$ for $\ket{g}$ and $210^\circ$ for $\ket{e}$ -- and faster noise rise away from it as entangler gain increases (Fig.~\ref{fig:TMS_noise}(a)). As shown in Fig.~\ref{fig:SNR}(a), optimal SNR is acheived with a relative pump phase of $190^\circ$ at which the total noise power for qubit in $\ket{g}$ and $\ket{e}$ states is minimized. At low entangler gain ($G_E \leq 2$~dB), the total noise power at this relative pump phase decreases with increasing entangler gain, leading to an increasing SNR. However, as noise reduction at the optimal pump phase saturates at higher enetangler gain, noise power at this relative pump phase would only increases, resulting in a decreasing SNR.

In the ideal case without photon loss and added noise from following amplifiers on the output line, the SNR of our system can be improved by as much as $190\%$ with entangler gain of 3~dB as shown in Fig.~\ref{fig:SNR}(b) (dash line). However, photon loss inside the interferometer limits the achievable SNR improvement to $72\%$, and noise added by following amplifiers on the output line further reduces it to $31\%$. Photon loss inside our system has been minimized by using components with the lowest loss that are commercially available, such as superconducting coaxial cables, and is on par with the state-of-art \cite{Flurin2012}. On the other hand, effects of photon loss and added noise on the output line, in principle, can be alleviated by operating the analyzer with higher gain. For example by operating analyzer with 30~dB gain, instead of 20~dB gain, such effects would be reduced by a factor of 10, resulting in an SNR improvement of $70\%$ instead of $31\%$ with entangler gain of 3 dB. However, the compression power and bandwidth of the analyzer would be reduced by a factor of 10 and $\sqrt{10}$ \cite{Frattini2018}, respectively. This method can be implemented in future experiments with amplifiers of higher compression power and larger bandwidth.

\subsection{Quantum efficiency of a TMSL measurement}

Another important figure-of-merit of a quantum measurement is its efficiency, which determines the fraction of information of the system being measured which is obtained by the observer, rather than  lost to all other potential observers\cite{Korotkov2016, Hatridge2013, Clerk2010, Murch2013}. Readout fidelity scales exponentially with quantum efficiency, and thus it plays a vital role in experiments which requires fast and high fidelity measurements, such as feedback control in quantum error correction \cite{Shankar2013}. 

In our experiment, the quantum efficiency $\eta$ of our qubit measurement is determined by analyzing its back-action on the qubit with a weak measurement protocol established in Ref.~\cite{Hatridge2013}. This protocol provides a self-calibrated way of determining the overall efficiency of a measurement system. As there is no well established theory for weak measurement back-action with our TMS interferometer, we focus our study on two special cases where it most closely resembles the coherent light measurement; the `TMS High' and `TMS Low' cases shown in Fig.~\ref{fig:Bull_eye}(a) where the output noise power is independent of the qubit state, and the interferometer's output resembles CS + PP readout with unusual noise values.   

In Fig.~\ref{fig:Quantum efficiency} we summarize the results of the back-action experiments with coherent light and two-mode squeezed light ($G_E = 0.5$~dB). Fig.~\ref{fig:Quantum efficiency}(a) shows the conditional  $x$ and $y$ components of the qubit state Bloch vector for measurement results that have zero in-phase component ($I = 0$) of coherent light readout. In CS + PP readout, the $x$- and $y$-component of the qubit receive a purely quantum stochastic back-action due to the knowledge gained by the experimenter from the $Q$-component of the readout pulse.  This effect can also be present in readout with coherent states plus phase-sensitive amplification (CS+PS). In CS+PS readout, the choice of amplified quadrature can shift from granting the observer only $I$-quadrature information (which we would usually call an optimal readout quadrature), and correspondingly maximal $z$-back action and no effect on the qubit's $x$- and $y$-components.  By rotating the phase of the amplifier by $\ang{90}$, we can reverse the situation, and produce only stochastic phase back-action\cite{Sliwa2016, Korotkov2016}.  For CS + PP readout, these two back-actions are equal in amplitude, as the amplifier favors neither quadrature. The amplitude and frequency of the $x$ and $y$-component oscillation, on the other hand, is determined by both the measurement strength and the efficiency of the measurement system. Therefore, by fitting this set of data, we can obtain the self-calibrated measurement strength and the overall efficiency of the measurement setup (see Supplemental for detailed procedure). From the data shown in Fig.~\ref{fig:Quantum efficiency}(a), we obtain a overall efficiency of $\eta=0.46$ for the coherent light readout method which sets the base line of our system. By accounting for the known loss in coherence due to qubit $T_2$, we can also calculate a corrected efficiency of $\eta_{cor}=0.52$.  

Figure~\ref{fig:Quantum efficiency}(b) and (c) show the results of applying the same protocol for two-mode squeezed light readout at the `TMS High' and `TMS Low' matched noise points, respectively. We have adjusted the drive amplitudes to match all three measurement conditions' SNR (and correspondingly their $z$-component back-action, see Fig.~\ref{fig:Tomo_Z}).   However, the data clearly show that both the oscillation frequency and amplitude of the $x$- and $y$- components are very different from the CS + PP case, which indicates that the back-action strength and quantum efficiency are very different. From the same fits, we obtain a quantum efficiency of 0.58 at the `TMS High' point, and 0.29 at the `TMS Low' point. 

The difference in the measurement efficiency with TMS light seems to be similar to the qubit readout with single-mode squeezed light and a phase-sensitive amplifier. For example, in Ref.~\cite{Eddins2018}, similar reduction in quantum efficiency was observed when noise in the information-carrying quadrature is squeezed, which was explained as a consequence of an effective reduction in the quantum efficiency of the output chain following the quantum-limited phase-sensitive amplifier; when quantum noise in the signal is reduced, the relative contribution of the fixed amount of noise added by the output chain is increased, thus the quantum efficiency of the output chain is effectively reduced, and so is the overall quantum efficiency. However, this possible cause can be ruled out in our experiment. As shown in Fig.~\ref{fig:NVR}, even if we vary the relative noise contributed by the analyzer and the output chain, the known changes in output noise cannot explain simultaneously the results at `TMS High' and `TMS Low' points. The disagreement between our data and the theoretical estimation suggests that an intuitive explanation is not sufficient. The full quantum explanation of the problem remains to be an interesting question in the quantum measurement theory to be addressed in future work.

\section {Conclusions and Outlook}
In this work, we have demonstrated a new scheme for dispersivem, interferometric readout of a superconducting qubit with two-mode squeezed light and phase preserving amplification. In this readout scheme, we can increase the SNR of projective readout by suppressing the noise output of our amplifier below the usual Cave's limit for a phase-preserving amplifier fed with unsqueezed vacuum. In our experiment, despite photon loss we have achieved a $31\%$ improvement in power SNR compared to conventional coherent light plus phase-preserving amplification readout. This improvement in SNR will result in a suppression of SNR-limited readout infidelity by a factor of 5, if one starts with a $1\%$ infidelity in coherent light readout, with even greater improvements at higher base fidelities. 

A still more interesting result emerges as we investigate the quantum readout efficiency of our TMS interferometer using weak measurements at points where the noise output is the same for both qubit states.  These data show that there are important effects on the ratio between $z$ back-action and the concomitant qubit phase back-action of this measurement process relative to other known readout schemes.  It appears that the increase in SNR at the `TMS Low' match point comes at the cost of reducing the trackability of phase back-action.  Conversely,  at the `TMS High' match point this phase trackability is enhanced.  Although we rule out post-interferometer noise as the source of this effect, more theoretical work is needed to understand the role of inefficiencies inside the interferometer due to imperfect squeezing/amplification and losses in the interferometer arms.  This limitation notwithstanding, the fact that the `TMS High' match point gives desirable quantum properties at a point which deliberately degrades SNR and fidelity of projective measurement should encourage exploration of measurement methods which are not just the quantum analogs of good classical measurement schemes. 

Finally, while tracking a single qubit's phase during measurement is not of direct value for single qubit measurements in quantum computing, measurement-based entanglement is a vital component of many error-correction schemes in quantum information, and in these schemes\cite{Silveri2016, Roch2014, Dickel2018}, maintaining/tracking two qubit phase coherence during a high-fidelity measurement is vital.  Our current experiment can be readily extended to two-qubit entanglement\cite{Silveri2016} by adding a second qubit on the lower arm, and we expect the ability to rebalance measurement strength and phase trackability {\it in-situ} to give crucial tolerance for losses and inefficiencies which currently limit these experiments.

\section{Acknowledgments}

We acknowledge helpful discussions with S. Barzanjeh, Chenxu Liu, and David Pekker. This work was supported by the Army Research Office under Grants No. W911NF15-1-0397 and W911NF-18-1-0144, by NSF Grant No. PIRE-1743717 and by the Charles E. Kaufman Foundation of the Pittsburgh Foundation. The views and conclusions contained in this document are those of the authors and should not be interpreted as representing the official policies, either expressed or implied, of the Army Research Office or the US Government. The US Government is authorized to reproduce and distribute reprints for government purposes notwithstanding any copyright notation herein.

\renewcommand{\theequation}{S.\arabic{equation}}
\renewcommand{\thefigure}{S.\arabic{figure}}
\setcounter{equation}{0}
\setcounter{figure}{0}
\section{Supplementary Information} \label{sec: SM}

\subsection{Cryogenic Microwave Setup}
\begin{figure*}
	\includegraphics[scale =0.8]{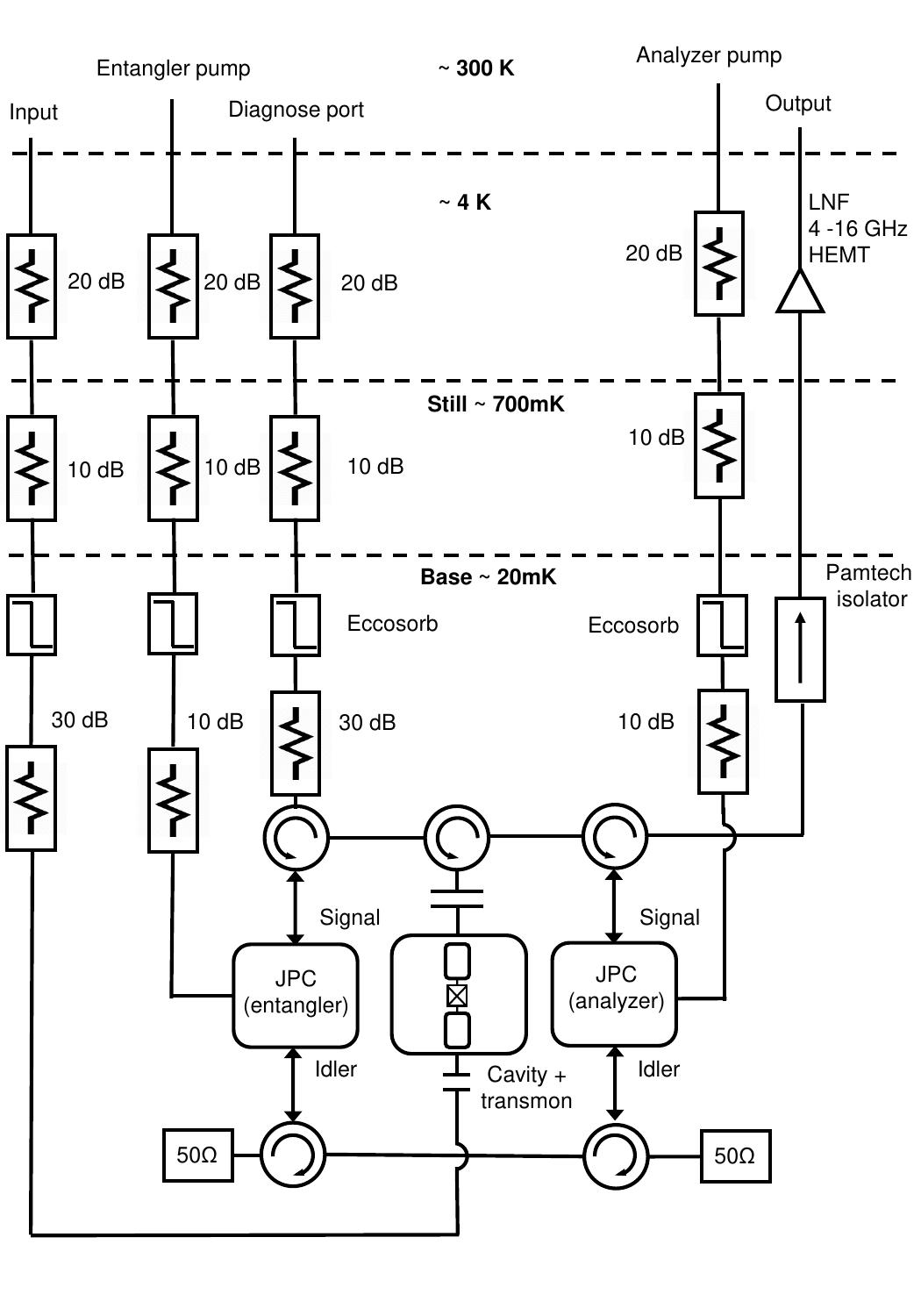}
		\caption{ \textbf{Wiring diagram of the cryogenic microwave measurement setup.} }
\label{fig:Microwave setup}
\end{figure*}

The experiment is cooled down to around 20 mK on the base stage of a dilution refrigerator. The cryogenic microwave setup is shown in Fig.~\ref{fig:Microwave setup}. 
Fridge input lines carrying signals from room temperature to the system are attenuated and filtered with homemade lossy Eccosorb filters. Room temperature electronics, which include microwave generators, IQ mixers and an arbitrary waveform generator, are used to produce microwave pulses to drive the qubit and cavity. Both the qubit and cavity drives are sent into the system through the qubit-cavity input line, which connects to the weak port of the cavity. The JPCs are pumped with Keysight microwave generator. The output signal from the TMS interferometer is amplified by a chain of low noise cryogenic and room temperature amplifiers before been down-converted, digitized and demodulated with a room-temperature reference copy. 

\subsection{Sample parameters} \label{sec:Exp_params}
The cavity in our experiment is a 3D aluminum coaxial post cavity with a resonance frequency of $\omega_{c} / 2 \pi = 7.447$~GHz, coupling quality factors of $Q_{strong}=752$ on the strong port, and $Q_{weak} \sim 1,000,000$ on the weak port. Therefore, the cavity linewidth seen from the strong port is $\kappa/2\pi = 9.9$~MHz. The superconducting qubit is a 3D transmon qubit made by commonly used Dolan bridge technique with ground to excited state transition frequency of $\omega_{ge} / 2 \pi = 4.102$~GHz, anharmonicity $\alpha = 180$~MHz, and a qubit-cavity dispersive coupling strength of $\chi / 2 \pi = 1.8$~MHz. This qubit has a $T_1$ of 18.2~$\mu s$, and $T_{2R}$ of $4.4~\mu s$ ($T_{2E}$ is $4.6~\mu s$). The cavity is placed in an aluminum shield that is inside a $\mu$-metal cryoperm shield. The whole system is wrapped in mylar.

\subsection{Pump leakage cancellation and characterization} \label{sec:pump_leakage}

Due the design of our JPC, applied pump tone preferentially leaves from the signal and idler ports. Similarly, a pump tone can enter a JPC through its signal and idler ports (though in practice we apply pump tones only to the designated pump port). Thus, in our experiment, a fraction of the pump signal of each JPC always leaks into the other JPC through the arms of the interferometer. As both JPCs are biased such that their mode frequencies are matched, this pump leakage could severely affect its operation. Thanks to the directionality of signal flow in the interferometer imposed by the circulators, pump leakage from analyzer JPC to entangler JPC is attenuated by -40~dB to -50~dB by the two circulators and can be safely neglected.

On the other hand, pump leakage from entangler JPC only experiences minimal attenuation on its path to analyzer JPC. Furthermore, the analyzer JPC is operated with high gain (20~dB) which makes it extremely sensitive to pump power variations; to the first order of pump power, a relative pump power variation of $\Delta p$ leads to a gain variation of  $\sqrt{G}(\sqrt{G}+1)^2 \Delta p$, namely a $1\%$ variation of the pump power at $G=100$ (20~dB) translates into a gain variation of $\Delta G = 10$ (0.4~dB). For example, with $G_E=4$~dB and $G_A=13$~dB at the interested frequency when they are turned on separately, we observe that $G_A$ varies between 6~dB and 31~dB depending on the relative pump phase when both amplifiers are on, which corresponds to a $\pm 50\%$ variation of the pump power in analyzer JPC. Therefore, to ensure proper operation of the interferometer, it is crucial to minimize the pump power variation in the analyzer caused by entangler pump leakage.

\begin{figure}
	\includegraphics[scale = 1.0]{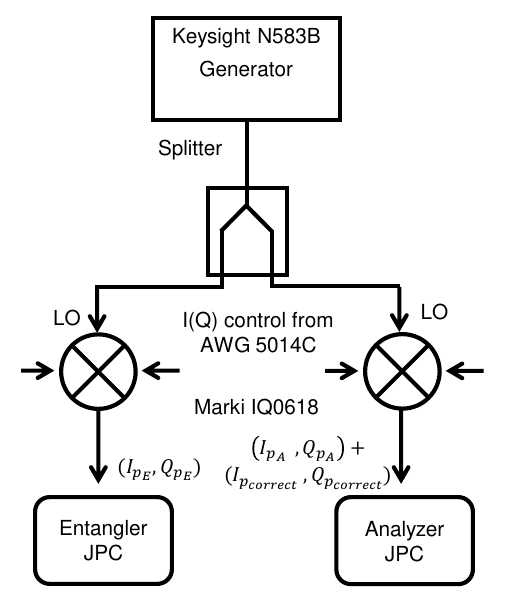}
		\caption{ \textbf{Wiring diagram of room temperature setup for canceling the pump leakage.} The output of a single generator (Keysight N583B) is split and fed to two I/Q mixers which control the two JPC pumps.  The Signal applied to the analyzer JPC is the sum of the desired pump and a correction designed to cancel leakage due to the entangler JPC pump.  The use of a single generator has the additional benefit of stabilizing the relative phase of the two pumps, as a drift in phase of the generator affects both pumps equally.}
\label{fig:pump leakage cancellation}
\end{figure}

\begin{figure}
	\includegraphics[scale = 1.0]{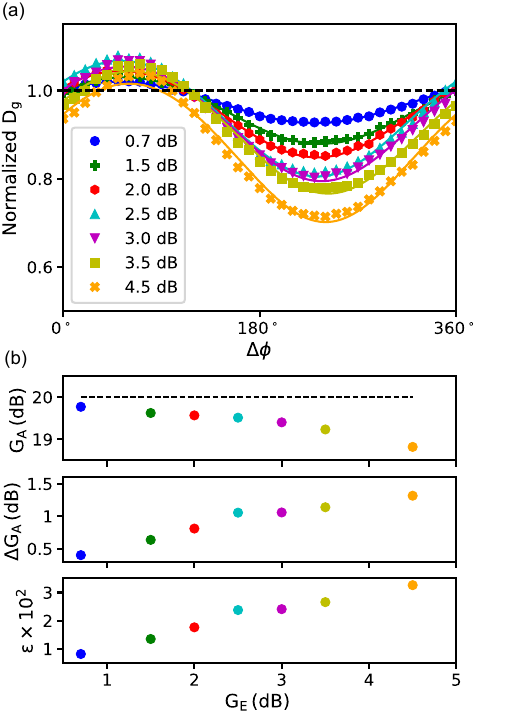}
		\caption{\textbf{Normalized signal strength of  qubit  readout  with  coherent  light and displaced two-mode squeezed vacuum for qubit in $\ket{g}$.} (a) Average output voltage of a displaced two-mode squeezed vacuum state normalized to that of a coherent light state versus relative pump phase for different entangler gains. Data are represented by colored symbols and the corresponding theoretical fittings are represented by solid lines of the same color. (b) Analyzer gain ($G_A$), gain variation ($\Delta G_A$) and relative amplitude of the entangler pump leakage ($\epsilon$) versus entangler gain from fitting data in (a) with Eq.~\ref{eq:D_norm}. Color of the symbols represent entangler gain.}
\label{fig:TMS pump leakage}
\end{figure}

To minimize the unwanted effect of entangler pump leakage, we deliver the pump signals to the JPCs through a circuit shown in Fig.~\ref{fig:TMS pump leakage} similar to the one used in Ref.~\cite{Flurin2012}. In this circuit, the entangler pump is split and fed to two I/Q mixers which provide both JPC pumps.  The analyzer's pump signal is a combination of a the desired analyzer pump and a phase and amplitude shifted copy of the entangler pump which cancels its leakage in the analyzer JPC.  Experimentally, we minimize the leakage by varying the analyzer pump phase and identifying the correction factor where analyzer gain is insensitive to the presence or absence of the entangler pump.  Another way of eliminating the pump leakage to the analyzer could be to add low-pass filters on both arms of the interferometer. However, this would introduce additional loss to the interferometer and degrade its performance. Therefore, we chose the active method described here. More, to prove that this effect is due entirely to pump leakage, and not two-mode squeezing effects in the interferometer, we verified the presence of pump leakage and our cancellation scheme with the up-stream, entangler JPC tuned far away in frequency with flux.

To quantify the amount of residual pump leakage in analyzer, we operate the analyzer at 20~dB gain pump condition and measure the output signal strengths when a coherent drive applied to the cavity weak port while sweeping the relative pump phase and entangler gain. Fig.~\ref{fig:TMS pump leakage}(a) shows the output signal strength normalized to that when the entangler is off for the qubit in $\ket{g}$. The oscillation of the output signal strength with relative pump phase manifests the interference between the residual entangler pump leakage and the analyzer pump. In this case, the total time-averaged pump power in analyzer can be written as
\begin{align}
    P = P_0(1+\epsilon \cos(\Delta \phi + \phi_0)+\epsilon^2)
\end{align}
where $P_0$ is pump power of the analyzer pump, $\epsilon$ is relative amplitude of the residual pump leakage, $\Delta \phi$ is the relative pump phaes, $\phi_0$ is the constant phase shift between the pump leakage and the analyzer pump. The term $\epsilon^2$ can be neglected in our analysis since $\epsilon \sim 0.01$.
Consequently, the gain of the analyzer, to the its first order dependence of pump power, can be written as
\begin{align}
    G^{\rm{TMS}}_A = G_A + \sqrt{G_A}(\sqrt{G_A}+1)^2 \epsilon \cos(\Delta \phi + \phi_0)
\end{align}
where $G_A$ is the gain of the analyzer without entangler pump leakage. The normalized output signal strength is 
\begin{align}
\label{eq:D_norm}
    \frac{D^{\rm{TMS}}}{D^{\rm{CS}}} = \sqrt{\frac{G_A + \sqrt{G_A}(\sqrt{G_A}+1)^2 \epsilon \cos(\Delta \phi + \phi_0)}{G^{\rm{CS}}_A}}
\end{align}
where $G^{\rm{CS}}_A=20$~dB is the analyzer gain when entangler is off. By fitting data shown in Fig.~\ref{fig:TMS pump leakage}(a) with Eq.~\ref{eq:D_norm}, we can get $G_A$ and $\epsilon$ for given entangler gain. In Fig.~\ref{fig:TMS pump leakage}(a) the fitting to the data of a given entangler gain is shown by solid lines of the same color. Fig.~\ref{fig:TMS pump leakage}(b) shows fitting results of $G_A$, $\epsilon$ and maximum gain variation $\Delta G_A = \epsilon \sqrt{G_A} (\sqrt{G_A}+1)^2$ with respect to relative pump phase for all the different entangler gains. 

The fitting gives an analyzer gain of $G_A=19.7$~dB with entangler on with gain $G_E \le 3$~dB. The small reduction (0.3~dB) of the analyzer gain with entangler on is a result of imperfect cancellation of the pump leakage. For entangler gain $G_E>3$~dB, the analyzer gain decreases as a consequence of gain compression caused by increasing power of the displaced TMS light. The residual entangler pump power leakage also increases with the entangler gain from 0.01 to 0.03 of the analyzer pump power, which is 40 to 13 times smaller than that without the cancellation circuit.

\subsection{Noise and SNR of qubit readout with TMS interferometer} \label{sec:TMS_thry}

The generation and detection of two-mode squeezed (TMS) vacuum with two non-degenerate phase-preserving Josephson parametric amplifier similar to JPC has been demonstrated in \cite{Flurin2012}. Here, we first summarize the relevant results from this work and then develop the theory for predicting the SNR improvement achievable in our experiment.

\subsubsection{Gain of TMS interferometer}
The input and output of the two modes, $a$ and $b$ or signal and idler, of a phase-preserving amplifier (e.g a JPC) is related by the scattering relations of the amplifier:
\begin{align}
    a_{out} &= S^\dag a_{in} S = \cosh{r} a_{in} + e^{i\phi}
    \sinh{r} b^\dag_{in} \\
    b^\dag_{out} &= S^\dag b^\dag_{in} S = e^{-i\phi}\sinh{r} a_{in} + \cosh{r} b^\dag_{in}
\end{align}
where $\cosh{r} = \sqrt{G}$ and $\sinh{r} = \sqrt{G-1}$ are the amplitude gain in reflection and in transmission, $\phi$ is the phase of the pump tone. When two identical JPCs are connected with their two ports respectively, it is straightforward to calculate the scattering parameters for the combined system using the equations above twice. For example, the output field operator of the signal port 2nd JPC -- the `analyzer', is
\begin{align}
        a_{A,out} =  S_{aa} a_{E,in} + S_{ab}^* b^\dagger_{E,in}
    \label{eq:a_out_ideal}
\end{align}
with $S_{aa}$ and $S_{ab}$ given by
\begin{align}
    S_{aa} = \cosh{r_E} \cosh{r_A} + e^{i\Delta\phi} \sinh{r_E} \sinh{r_A}
    \label{eq:Saa_ideal} \\
    S_{ab} = e^{-i\phi_E}(\cosh{r_A} \sinh{r_E} + e^{-i\Delta\phi} \sinh{r_A} \cosh{r_E})
    \label{eq:Sab_ideal}
\end{align}
where $\phi_E$, $\phi_A$ are the pump phase of the two amplifiers, and $\Delta \phi = \phi_A - \phi_E$ is the relative pump phase between them. $S_{aa}$ and $S_{ab}$ represent the amplitude gain for input signal to the signal and idler port of the first JPC, respectively. It's straight forward to see that the amplitude of $S_{aa}$ and $S_{ab}$ will both vary with the relative pump phase $\Delta\phi$. Especially, when two JPCs have a matched gain ($r_E = r_A = r$) and a relative pump phase of $\pi$, the total gain of the system will become $\cosh^2{r} - \sinh^2{r} = 1$, indicating that the output signal power will be the same as the input signal. In the ideal case, when there is no loss and no added noise inside the interferometer, the output noise will just be the uncorrelated vacuum noise present at the input of the interferometer.

\begin{figure}
	\includegraphics[scale = 1.0]{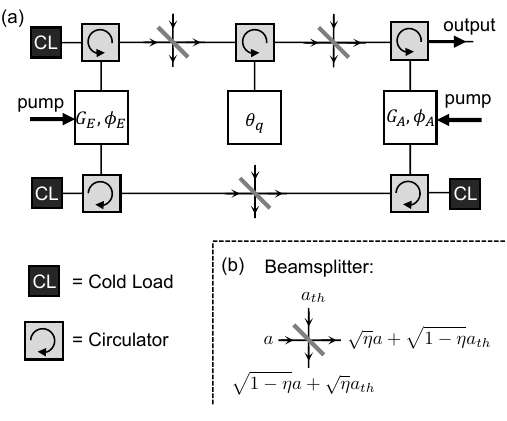}
		\caption{\textbf{Two-mode squeezed light interferometer with beamsplitters respresenting loss in the signal pathway}. (a) Entangler and analyzer JPCs are represented by squares with their gain and pump phase, respectively. Square with symbol $\theta_q$ represents the qubit-cavity system which introduces an extra qubit state dependent phase shift of $\theta_q$ to light propagating on the upper arm of the interferometer. (b) Photon loss is modeled by a beamsplitter with power transmission coefficient $\eta$. It transforms the signal mode operators $a$ and bath mode operator $a_{th}$ of the same frequency into $\sqrt{\eta}a + \sqrt{1-\eta}a_{th}$ and $\sqrt{1-\eta}a + \sqrt{\eta}a_{th}$ in transmission, respectively.}
\label{fig:beamsplitter_model}
\end{figure}

In practice, there is always photon loss inside amplifiers and along the paths between the two amplifiers. Without distinguishing these different loss mechanism, we model photon loss in our TMS interterferometer with beamsplitters which partially transmit the incoming signal and couples uncorrelated noise from background thermal baths into the interferometer \cite{Flurin2012}.

Fig.~\ref{fig:beamsplitter_model}(a) shows a schematic of our model of the interferometer. In order to accurately capture the effect of photon loss inside the TMS interferometer on qubit readout with TMS light, we use two beamsplitters -- before and after the qubit-cavity system -- to model loss in the upper arm. Due to the symmetry of the upper arm with respect to the qubit cavity system, the two beamsplitters are identical. The power transmission coefficient of the beamsplitters on the upper arm is $\eta_a$ and that of the beamsplitter on the lower arm is $\eta_b$. The qubit-cavity system is modeled as a phase shifter which adds a qubit-state-dependent phase shift $\theta_q$ on the upper arm where $\theta_q = \theta$ for qubit in $\ket{g}$, and $-\theta$ for qubit in $\ket{e}$. The output at the signal port of the analyzer now becomes 
\begin{align}
    a_{A,out} =  S_{aa} a_{E,in} + S_{ab}^* b^\dagger_{E,in} + M_{aa} a_{th} + M_{ab}^* b^\dagger_{th}
    \label{eq:a_out_w_loss}
\end{align}
with $S_{aa}$, $S_{ab}$, $M_{aa}$ and $M_{ab}$ given by 
\begin{widetext}
    \begin{align}
    S_{aa} &= e^{i\Delta \phi} \left( e^{i(\theta_q - \Delta \phi)} \eta_a \cosh{r_A} \cosh{r_E} + \sqrt{\eta_b} \sinh{r_A}\sinh{r_E}\right) 
    \label{eq:Saa_w_loss}\\
    S_{ab} &= e^{-i\phi_A} \left( e^{-i(\theta_q- \Delta \phi)} \eta_a \cosh{r_A} \sinh{r_E} + \sqrt{\eta_b} \sinh{r_A}\cosh{r_E} \right)
    \label{eq:Sab_w_loss} \\
    M_{aa} &= (1+\sqrt{\eta_a}e^{i\theta_q}) \sqrt{1-\eta_a} \cosh{r_A}
    \label{eq:Maa}\\
    M_{ab} &= e^{-i\phi_A} \sqrt{1-\eta_b} \sinh{r_A}
    \label{eq:Mab}
    \end{align}
\end{widetext}
Comparing to Eq.~\ref{eq:a_out_ideal}, the two extra terms in Eq.~\ref{eq:a_out_w_loss} represent the contribution of noise to the output signal. Comparing Eqs.~\ref{eq:Saa_ideal} - \ref{eq:Sab_ideal} with Eqs.~\ref{eq:Saa_w_loss} - \ref{eq:Sab_w_loss}, we see the loss in the interferometer indeed leads to reduction of gain of the input signal. 

The qubit-cavity system, on the other hand, introduces a qubit-state-dependent phase shift between the two arms through which the qubit state inforamtion is encoded in the gain of the interferometer. If the qubit measurement singal enters the interferometer from the input port the entangler \cite{Barzanjeh2014}, then both its average amplitude and fluctuation measured at the output of the analyzer would carry the qubit state information. However, this way of measuring qubit state with TMS light has the drawback that the average amplitude of the signal is also reduced when the noise is reduced. This effectively reduces SNR of the signal at room temperature since a fixed amount of noise will be added to it by the low noise amplifiers (LNAs) on the output line. In our experiment, the qubit measurement signal entering from the weak port of the cavity will only interact with the analyzer, therefore the average amplitude of the output signal will be independent of the qubit state. Instead, the fluctuation/noise of the output signal will carry the qubit state information.         

\subsubsection{Noise of TMS interferometer}
In experiment, we measure both the in-phase ($I$) and quadrature ($Q$) components of the output signal, which are related to the field operators by
\begin{equation}
    I = \frac{a+a^\dagger}{2}, \quad Q = i \frac{a^\dagger-a}{2}.
\end{equation}
The noise power of the output signal is proportional to the variance of the signal $\sigma^2$, which is given by
\begin{align}
    \sigma^2 = \sigma^2_I + \sigma^2_Q 
\end{align}
with
\begin{align}
    \sigma^2_{I} & = \expval{I^2} - \expval{I}^2 \nonumber \\
    & = \frac{1}{4}\left( \expval{a^2} - \expval{a}^2 + \expval{a^{\dagger 2}} - \expval{a^\dagger}^2 \right) \nonumber \\
    & \quad + \frac{1}{2} \left(\expval{a^\dagger a} - \expval{a^\dagger} \expval{a} \right) + \frac{1}{4} \\
    \sigma^2_{Q} & = \expval{Q^2} - \expval{Q}^2 \nonumber \\
    & = - \frac{1}{4}\left( \expval{a^2} - \expval{a}^2 + \expval{a^{\dagger 2}} - \expval{a^\dagger}^2 \right) \nonumber \\
    & \quad + \frac{1}{2} \left(\expval{a^\dagger a} - \expval{a^\dagger} \expval{a} \right) + \frac{1}{4}
\end{align}
Under the condition that both the input fields of the entangler and the bath modes are in vacuum states, for the output of the signal mode of the analyzer we have
\begin{align}
    \sigma^2_{I,out}, \sigma^2_{Q,out}  & = \frac{1}{2}\expval{a_{A,out}^\dagger a_{A,out}} + \frac{1}{4}    
\end{align}

From Eq.~\ref{eq:a_out_w_loss}, we get
\begin{widetext}
\begin{align}
    \expval{a_{A,out}^\dagger a_{A,out}} & = \expval{(\alpha^*_1 a^\dagger_{E,in} + \alpha_2 b_{E,in} + \alpha^*_3 a^\dagger_{th} + \alpha_4 b_{th})(\alpha_1 a_{E,in} + \alpha^*_2 b^\dagger_{E,in} + \alpha_3 a_{th} + \alpha_4^* b^\dagger_{th})} \\
    & = |\alpha_2|^2 \expval{b_{E,in}b^\dagger_{E,in}} + |\alpha_4|^2 \expval{b_{th}b^\dagger_{th}} \\
    & = |e^{i(\theta_q- \Delta \phi)} \eta_a \cosh{r_A} \sinh{r_E} + \sqrt{\eta_b} \sinh{r_A}\cosh{r_E}|^2 + (1-\eta_b)\sinh^2{r_A} \\
    & = (\eta^2_a \cosh^2{r_A} + \eta_b \sinh^2{r_A}) \sinh^2{r_E} + \frac{\eta_a\sqrt{\eta_b}}{2} \sinh{2r_A} \sinh{2r_E} \cos{(\theta_q - \Delta \phi)} + \sinh^2{r_A}
\end{align}
\end{widetext}
Therefore, 
\begin{widetext}
\begin{align}
    \sigma^2_{I,out}, \sigma^2_{Q,out} & = (\eta^2_a \cosh^2{r_A} + \eta_b \sinh^2{r_A}) \frac{\sinh^2{r_E}}{2} + \frac{\eta_a\sqrt{\eta_b}}{4} \sinh{2r_A} \sinh{2r_E} \cos{(\theta_q - \Delta \phi)} + \frac{\cosh{2r_A}}{4} 
    \label{eq:TMS_noise_output}
\end{align}
\end{widetext}
In the special case of $r_E=0$, namely entangler been off, the output noise photon number is $\cosh{2r_A}/2$, which is the output noise of a quantum-limited phase-preserving amplifier with vacuum noise input on its signal and idler port.

Equation~\ref{eq:TMS_noise_output} gives the output noise photon number at the signal port of the analyzer. It is related to the noise photon number measured at room temperature by
\begin{align}
    \sigma^2 = G_{\rm{out}}(\sigma^2_{out} + N_{out})
    \label{eq:TMS_noise_output_RT}
\end{align}
where $G_{out}$ is the total gain of the output line, $N_{out}$ is the added noise photon number referred back to the input of the HEMT. Therefore, the ratio of TMS light noise output and coherent light noise output measured at room temperature is
\begin{align}
    \frac{\sigma^2}{\sigma^2_{\rm{CS}}} = \frac{\sigma^2_{out} + N_{out}}{\cosh{2r_A}/2 + N_{out}}
    \label{eq:TMS_noise_norm}
\end{align}

\subsubsection{SNR of qubit readout with TMS interferometer}
As defined in Eq.~\ref{eq:SNR}, the SNR of qubit readout is 
\begin{equation}
{\rm{SNR}} = \frac{(I_c^g - I_c^e)^2 + (Q_c^g - Q_c^e)^2}{\sigma_g^2 + \sigma_e^2}
\end{equation}
where the signal strength is given by the separation between the distributions of measurement results corresponding to qubit in $\ket{g}$ and $\ket{e}$. For a given qubit-cavity system, this separation is proportional to the strength of the measurement pulse and total amount of amplification. In our experiment, the strength of the measurement pulses are the same for TMS light and coherent light readout, and the total gain of the LNAs following the analyzer is constant, therefore, the ratio between SNR of TMS light readout and coherent light readout recorded at room temperature is
\begin{align}
    \frac{\rm{SNR}}{\rm{SNR_{CS}}} = \frac{\sigma^2_{g} + \sigma^2_{e}}{\sigma^2_{\rm{CS},g} + \sigma^2_{\rm{CS},e}} \frac{G^{\rm{TMS}}_A}{G^{\rm{CS}}_A}
    \label{eq:TMS_SNR_norm}
\end{align}
where analyzer gain variation due to residual entangler pump leakage has been taken into account.

\subsubsection{Photon loss inside TMS interferometer}

Eq.~\ref{eq:TMS_noise_norm} and Eq.~\ref{eq:TMS_SNR_norm} give the expressions for noise and qubit readout SNR of TMS light normalized to that of the coherent light. In order to compare their predictions to data, we need to know the following parameters: $\eta_a$, $\eta_b$ and $N_{\rm{out}}$. In this section, we show how to determined these pareameters in our experiment.

First, the ratio $\eta_a^2\eta_b$ is determined by measuring $S_{aa}$ (\ref{eq:Saa_w_loss}) with a Vector Network Analyzer (VNA). In particular, $S_{aa}$ is measured under four different conditions: (1) both entangler and analyzer are off, thus $S_{aa} = e^{i\theta_q}\eta_a$; (2) entangler on and analyzer off, thus $S_{aa} = e^{i\theta_q}\eta_a \cosh{r_E}$; (3) entangler off and analyzer on, thus $S_{aa} = e^{i\theta_q}\eta_a \cosh{r_A}$; and (4) entangler and analyzer both on, thus $S_{aa}$ takes the form in Eq.~\ref{eq:Saa_w_loss}. 

Although it seems straight forward to determine $\eta_a$ by measuring $S_{aa}$ with both entangler and analyzer off, the measurement result recorded at room temperature is $G_{\rm{sys}}S_{aa}$ rather than $S_{aa}$, where $G_{\rm{sys}}$ is the total gain of the system which is determined by the total attenuation on the input line and total gain on the output line. To determine $G_{\rm{sys}}$, separate measurements are needed. Therefore, a measurement under condition (1) alone can not determine the value of $\eta_a$. However, it can be used to normalize the measurement results under conditions (2) and (3) to get gain of the entangler ($\cosh^2{r_e}$) and analyzer gain ($\cosh^2{r_A}$), respectively. By combining these results with that of measurement under conditions (4), we can determine $\eta_a^2\eta_b$.

We measured $S_{aa}$ for a range of entangler gain $G_E = \cosh^2{r_E}$ and pump phase difference $\Delta \phi$ while keeping the analyzer gain to $G_A = \cosh^2{r_A} = 10$~dB and qubit in state $\ket{g}$. We perform this characterization measurement with relatively small analyzer gain (10~dB) to minimize the effect of gain variation due to entangler pump leakage, and unwanted higher-order effects in the analyzer. 

\begin{figure}
	\includegraphics[scale = 1]{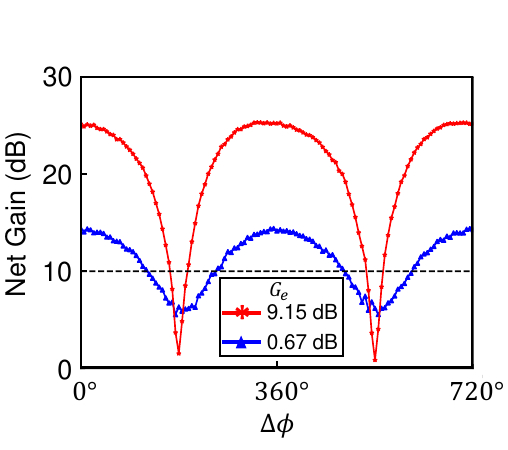}
		\caption{\textbf{Transmission measurement of the S parameter $S_{aa}$ of the TMS interferometer as a function of the relative pump phase}. The gain of the analyzer JPC, $G_A$, is fixed to be $10$~dB during the measurement (to avoid gain saturation for large entangler gain settings) as shown by the black dashed line. The blue triangle and red star trace corresponds to the different gain of the entangler JPC.} 
\label{fig:gain vs pump phase}
\end{figure}

As examples, Fig~\ref{fig:gain vs pump phase} shows data for $G_E = 0.67$~dB and 9.15~dB. By fitting these data with Eq.~\ref{eq:Saa_w_loss}, we can extract the ratio of the transmission coefficient of the upper and lower arms, $\eta^2_a/\eta_b$, which is found to be 0.8. This imbalance in transmission between the two arms is mostly due to the insertion loss of the extra circulator and cables on the upper arm that connects to the qubit-cavity system. Typical insertion loss of this circulator (Qunistar 4 to 8~GHz cryogenic circulator) is 0.2 to 0.5~dB per pass, which would give 0.4 to 1.0~dB (0.1 to 0.2) of additional loss on the upper arm. We estimate that extra cables in the upper arm contributes to about 0.5~dB loss to the signal. 

In order to determine the absolute value of $\eta_a$ and $\eta_b$, in principle we would need to know $G_{\rm{sys}}$ -- the total gain of the system -- as discussed earlier. However, as the power transmission coefficients also represent the quantum efficiency of transmitting information with photons along these signal pathways, we can deduce their values from the quantum efficiency of the system. 

In this case, we determine the quantum efficiency, $\eta$, of the part of system between the qubit-cavity and the room temperature electronics by performing weak measurements on the qubit with coherent light. It is related to $\eta_a$ by
\begin{align}
    \eta = \eta_a \eta_{\rm{out}}
    \label{eq:eta_sys}
\end{align}
where $\eta_{\rm{out}}$ is the quantum efficiency of the output line, which is dominated by the quantum efficiency of HEMT. From the weak measurement (next section), we find the $\eta =0.52$ (0.46 without correction of qubit $T_2$ effect). 

The quantum efficiency of the output line is directly related to the added noise photon number of the output line $N_{out}$ by,
\begin{align}
    \eta_{out} = \frac{\sigma^2}{\sigma^2+N_{out}}
    \label{eq:eta_out}
\end{align}
where $\sigma^2$ is the output noise photon number of the analyzer. The added noise photon number of the output line is determined from noise visibility ratio (NVR) of the system which is defined as
\begin{align}
    NVR = \frac{\sigma^2+N_{out}}{1/2 + N_{out}}
    \label{eq:NVR}
\end{align}
the ratio between output noise power with analyzer on and that with analyzer off while only vacuum noise is present at its input. 

In our experiment, the NVR is measured to be 7~dB when the analyzer gain is 20~dB. Under the assumption that analyzer is quantum-limited, we get $N_{out}=24.8$ which corresponds to a noise temperature of 8.9~K at 7.5~GHz which is typical for such a system. From Eq.~\ref{eq:eta_out}, we then get $\eta_{out} = 0.8$. Combining these results with Eq.~\ref{eq:eta_sys}, we get $\eta_a = 0.65$. From $\eta^2_a/\eta_b = 0.8$, we get $\eta_b = 0.53$.  

\begin{figure}
	\includegraphics[scale = 1]{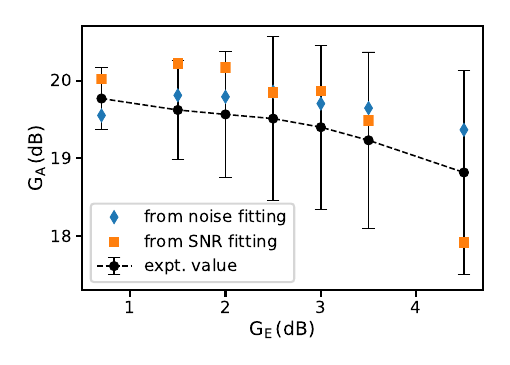}
		\caption{\textbf{Fitting value of analyzer gain ($G_A$)}. Analyzer gain ($G_A$) obtained from fitting the TMS noise data (Fig.~\ref{fig:TMS_noise}(a)) and the SNR data (Fig.~\ref{fig:SNR}(a)) is shown by blue diamond and orange square, respectively. Black dots represent analyzer gain shown determined from Fig.~\ref{fig:TMS pump leakage}(a). Error bar for each data point is given by $\Delta G_A$ in Fig.~\ref{fig:TMS pump leakage}(b)}.  
\label{fig:fitted_Ga_all}
\end{figure}

With all these parameters determined, we can now directly calculate the normalized output noise and qubit readout SNR and compare them with experiment data. However, as mentioned earlier that analyzer gain varies slightly with the relative pump phase due to entangler pump leakage, so we instead use Eq.~\ref{eq:TMS_noise_norm} and Eq.~\ref{eq:TMS_SNR_norm} to fit the normalized output noise and qubit readout SNR with the analyzer gain as the only fitting parameter, respectively. The fitting curves are shown in Fig.~\ref{fig:TMS_noise}(a) and Fig.~\ref{fig:SNR}(a) with colors corresponding to their respective data sets. Optimal noise suppression and SNR extracted from the data (color dots) and calculated from Eq.~\ref{eq:TMS_noise_norm} and Eq.~\ref{eq:TMS_SNR_norm} are shown in Fig.~\ref{fig:TMS_noise}(b) and Fig.~\ref{fig:SNR}(b), respectively. Analyzer gain obtained from these fittings compared to that obtained from the pump leakage fitting in Fig.~\ref{fig:fitted_Ga_all}. Except at $G_E=4.5$~dB, analyzer gain values from the noise fitting and SNR fitting agrees reasonably well with each other, and are both with in the range of gain variation caused by entangler pump leakage. Both fittings give average analyzer gains over the full range of relative pump phase which are expected to be different from the gain with entangler off (black dots). In addition, analyzer gain obtained from fitting the SNR data is an average of the values for qubit in $\ket{g}$ and $\ket{e}$ which are slight different from each other. Therefore, analyzer gain from noise fitting shows better overall agreement with the experiment value than that from SNR fitting.

\subsection{Back action of weak measurement}

\begin{figure}
	\includegraphics[scale = 1.0]{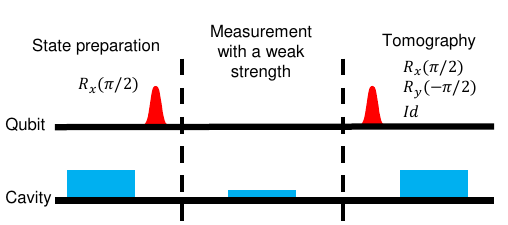}
		\caption{\textbf{Pulse sequence for quantifying measurement back action.} This pulse sequence consists three stages of qubit and cavity manipulation. The first stage is state preparation, during which the qubit is first projected to the state $\ket{g}$ by a strong measurement (blue box) and post-selection, and then rotated to the state $(\ket{g}+i\ket{e})/\sqrt{2}$ by a $R_x(\pi/2)$ pulse (red Gaussian). The second stage is weak measurement, during which the qubit state is measured by a weak measurement pulse (thin blue box) and an outcome $(I_m, Q_m)$ is recorded. In the third and final stage a qubit state tomography is performed by a combination of qubit rotation pulses (red Gaussian) $R_x(\pi/2)$, $R_y(-\pi/2)$ or $Id$ (no pulse) and a strong measurement (blue box) pulse. Note that a similar pulse sequence is used to record the back-action of the noise measurements, see Fig.~\ref{fig:Bull_eye}. The only different in the noise measurement is to replace the middle weak measurement with a record of noise of the system. That is, we apply no cavity drives, but only digitize the output noise of the system.}
\label{fig:Pulse_sequence}
\end{figure}

\begin{figure*}
\centering
	\includegraphics[scale = 0.55]{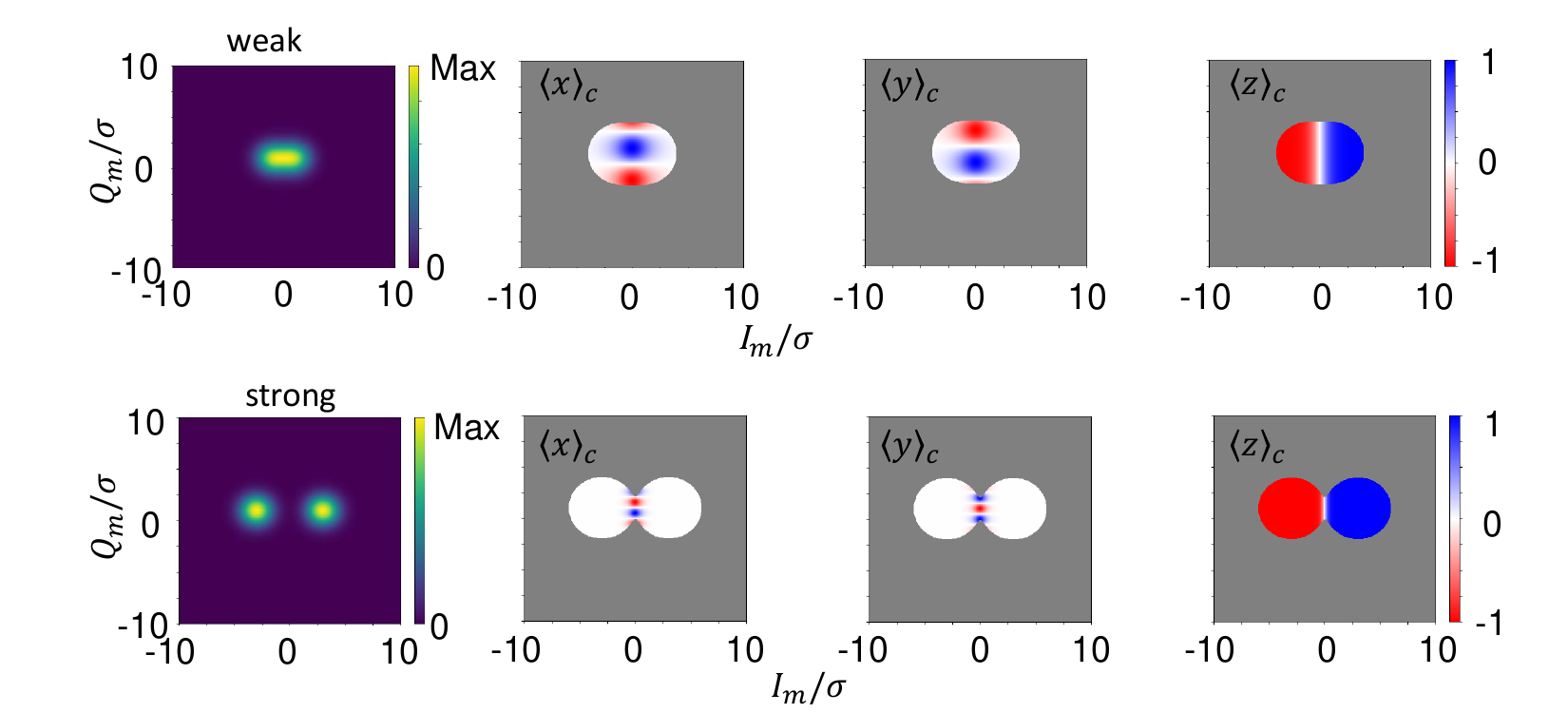}
		\caption[\textbf{Theory result for the measurement back-action}]{\textbf{Theory result for the measurement back-action.} Theory results for the back-action of a `weak' ($\bar{I}_m/\sigma = 1.0$) and `strong' ($\bar{I}_m/\sigma = 3.0$) measurement. The three histograms are showing $\langle X \rangle_c$, $\langle Y \rangle_c$ and $\langle Z \rangle_c$ versus the associated ($I_m/\sigma$, $Q_m/\sigma$) bin.}
\label{fig:weak measurement result}
\end{figure*}

\begin{figure*}
	\includegraphics[scale = 1.0]{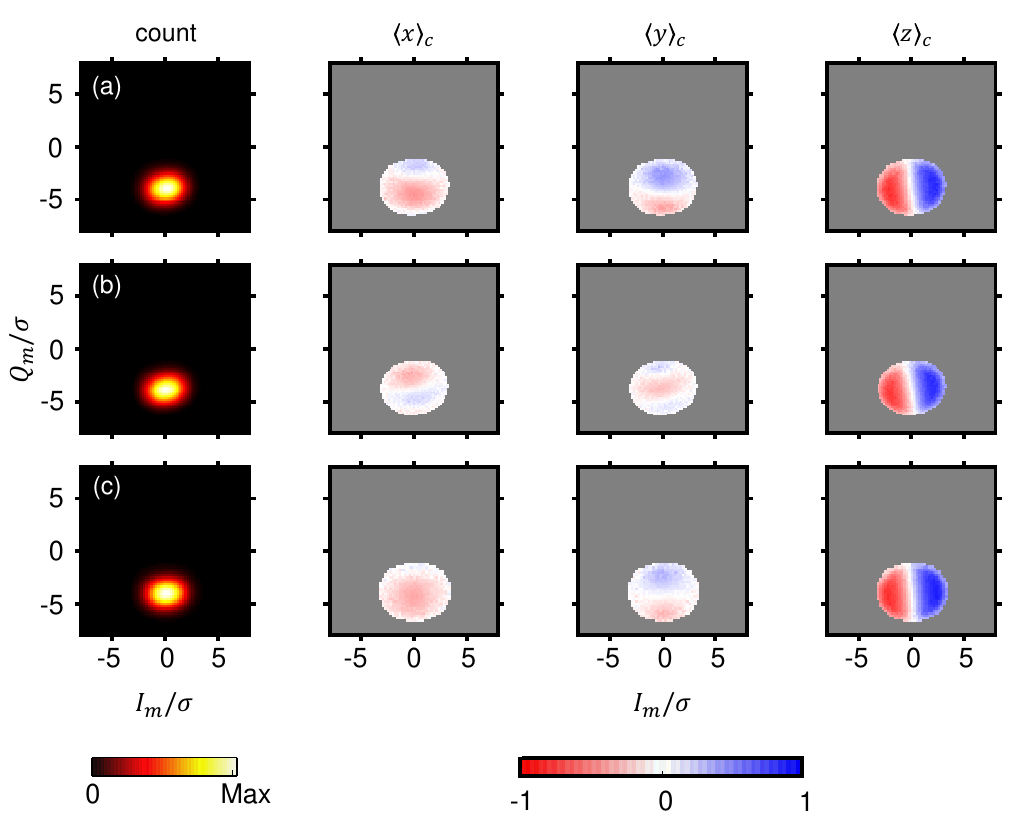}
		\caption{ \textbf{Experiment data for back action of weak measurement.} Results are shown for the back action of the weak measurement for coherent light (a) and two-mode squeezed light at high(b) and low(c) noise level match point. In all three cases, the strength of the weak measurement is $\bar{I}_m/\sigma = 0.66$. The leftmost  column shows the 2D histograms of scaled measurement outcomes recorded during the weak measurements. The right three columns are the conditional tomography data for $\expval{x}$, $\expval{y}$ and $\expval{z}$ component versus the associated $(I_m/\sigma, Q_m/\sigma)$ bin. The value in each bin is the average of all tomography data associated with that $(I_m, Q_m)$ value.}
\label{fig:Science protocol}
\end{figure*}

In this section we will give a brief introduction of the weak measurement protocol that we used to calibrate the quantum efficiency of our measurement. 
A quantum measurement will perturb the state of the qubit.
This perturbation is known as the back-action of the measurement. 
For a measurement performed with ideal quantum hardware, the qubit will still be in pure state after the measurement, as 
the evolution of the state can be perfectly tracked from the measurement record.  
In the usual coherent light measurement scheme, in order to measure a qubit in the state $\ket{\psi} = c_g \ket{g} + c_e \ket{e}$, we send a coherent light probe signal to the qubit-cavity system. 
After the probe signal traverses the cavity, the coherent light and the qubit form a new entangled state together: $\ket{\Psi} = c_g \ket{g} \ket{\alpha_g} + c_e \ket{e} \ket{\alpha_e}$. 
Here $\ket{\alpha_{g, e}}$ is the transmitted coherent state when the qubit is the ground or excited state, respectively. 
This new entangled state will serve as the pointer state of the system and will be sent through a quantum-limited amplifier (e.g a JPC) so the signal can be received and processed at room temperature.  

The effect of the back-action of a phase-preserving measurement can be understood as follows:
each shot of outcome will contain the measurements of both quadrature of the output mode, which we denote as $(I_m, Q_m)$. 
This I-Q pair will be used to determine the state of the qubit after the measurement and as been pointed out in Ref~\cite{Hatridge2013, Korotkov2016}, it contains all the information to track the state evolution of the qubit (provided the measurement is lossless).
That is, we can reconstruct the Bloch vector for the qubit after the measurement given the measurement result $(I_m, Q_m)$. 
However, in an actual experiment, there will also be losses.
Therefore, we introduce the concept of measurement efficiency $\eta = \frac{\sigma_{Heis}^2}{\sigma_{Heis}^2 + \sigma_{add}^2}$ as a way to measure how much quantum information is collected by the observer. 
Here $\sigma_{Heis}$ is the fundamental quantum noise from Heisenberg's uncertainty principle and the $\sigma_{add}$ is the noise added by the amplification chain in the measurement setup. 
As an example, for a qubit that is initially prepared in the state $\ket{+y} = (\ket{g} + \ket{e})/\sqrt{2}$, the final qubit Bloch vector $(x_f, y_f, z_f)$ is a function of measurement outcome $(I_m, Q_m)$(for simplicity we will neglect the qubit decoherence and the all losses before the quantum-limited amplifier).  Hatridge \textit{et al.} ~\cite{Hatridge2013} find: 
\begin{align}
    x_f(I_m, Q_m) &= \text{sech}(\frac{I_m \bar{I}_m}{\sigma^2}) \cross \nonumber \\
    & \text{sin}[\frac{Q_m \bar{I}_m}{\sigma^2} + \frac{\bar{Q}_m \bar{I}_m}{\sigma^2} (\frac{1-\eta}{\eta})] e^{-\frac{\bar{I}^2_m}{\sigma^2} (\frac{1-\eta}{\eta})} \label{Eq:backactionX} \\
    y_f(I_m, Q_m) &= \text{sech}(\frac{I_m \bar{I}_m}{\sigma^2}) \cross \nonumber \\
    &\text{cos}[\frac{Q_m \bar{I}_m}{\sigma^2} + \frac{\bar{Q}_m \bar{I}_m}{\sigma^2} (\frac{1-\eta}{\eta})] e^{-\frac{\bar{I}^2_m}{\sigma^2} (\frac{1-\eta}{\eta})} \label{Eq:backactionY}\\
    z_f(I_m, Q_m) &= \text{tanh}(\frac{I_m \bar{I}_m}{\sigma^2}) \label{Eq:backactionZ} 
\end{align}
where $\bar{I}_m$, $\bar{Q}_m$ and $\sigma$ are the center and standard deviation of 2D Gaussian distribution corresponding to the coherent measurement pulse. 
The above results indicate that by observing the measurement back-action, we can obtain the measurement efficiency as a parameter to assess the quality of the measurement apparatus. 

We use the pulse sequence shown in Fig.~\ref{fig:Pulse_sequence} to obtain the measurement back-action. 
We first use a strong projective measurement and record the data from which we will post-select to pick out the runs with qubit found in the  ground state. 
Next, we rotate the qubit into the $\ket{+y}$ state with a $\pi/2$ pulse. 
Then, we deliberately apply a measurement with a strength weaker than the projective measurement, and record the result. 
Finally, we do  full tomography on the qubit state to determine its Bloch components with another projective measurement. 
To get the back-action result for a certain weak measurement outcome $(I_m, Q_m)$, the Bloch components of the qubit $(\langle x \rangle_c, \langle y \rangle_c, \langle z \rangle_c)$ is calculated from the tomography result conditioned recording that specific  weak measurement result in a finely-binned histogram. 

A theoretical result is shown in Fig.~\ref{fig:weak measurement result}. 
When the measurement strength is weak, the ground and excited state distribution largely overlap with each other. 
Their separation will increases as the measurement strength (or SNR) increases. 
For the strong measurement case, by definition the two distributions are well separated. 
The conditional tomography data $\langle X \rangle_c$, $\langle Y \rangle_c$ and $\langle Z \rangle_c$ versus measurement outcome ($I_m/\sigma$, $Q_m/\sigma$) is also shown in the figure. 
When the measure strength is weak, the qubit state receives minimal perturbation from the measurement, with the outcomes corresponding most probably to Bloch vectors along the $+y$ direction. In practice, any outcome is possible, but outcomes which are very improbable require too many measurements to be practically observed, and so we truncate the theory and set all points to gray which are more than XX \gl{??} $\sigma$ away from the centers of the two Gaussian blobs.
We can also see the gradients in $\langle X \rangle_c$ along the $Q_m$ axis and $\langle Z \rangle_c$ along the $I_m$ axis, indicating the back-action of the measurement.  In an actual experiment, after getting these back-action data with this weak measurement protocol, by fitting the x- and y- conditional tomography data for outcomes where the z-component is zero (here $I_m=0$ to the theory, we can extract the measurement efficiency as a single fit parameter by comparing the frequency of oscillation (which we can extract from the histogram) to the amplitude of oscillations using the theory expression above.

When the measurement is strong, the qubit is then projected to $+z$ direction for positive $I_m$ values and $-z$ direction for negative $I_m$ values. The oscillations in $\langle X \rangle_c$  and  $\langle Y \rangle_c$ (where the $z$ back-action is weak), are still visible very near $I_m=0$ but are both very improbable (hence nearly being pinched off by our chosen cutoff in the graphs) and very quickly varying which will be almost impossible to detect in a practical lossy environment with a reasonable number of repetitions of the protocol.  Note that we have set a relatively `weak' projective measurement to make this effect more visible.

Fig.~\ref{fig:Science protocol}(a) shows the data of the qubit state tomography for weak measurements with coherent light (top row). This result is similar to what we get from the theoretical model in Fig.~\ref{fig:weak measurement result}(b) but with less contrast which is due to the finite measurement efficiency of the system. 
Data shown in Fig.~\ref{fig:Quantum efficiency} are line cuts of the $x$ and $y$ tomography data along $I_m = 0$, where there is no $z$ back-action. Oscillation in $\expval{x}_c$ and $\expval{y}_c$ shows the stochastic phase back action.
By fitting the back-action data of a weak measurement to the theory model given in Eq.~\ref{Eq:backactionX}, \ref{Eq:backactionY} and \ref{Eq:backactionZ}, we obtained the overall efficiency for the measurement reported in the paper by comparing the frequency and amplitude of the oscillations.

\begin{figure}
	\includegraphics[scale = 1.0]{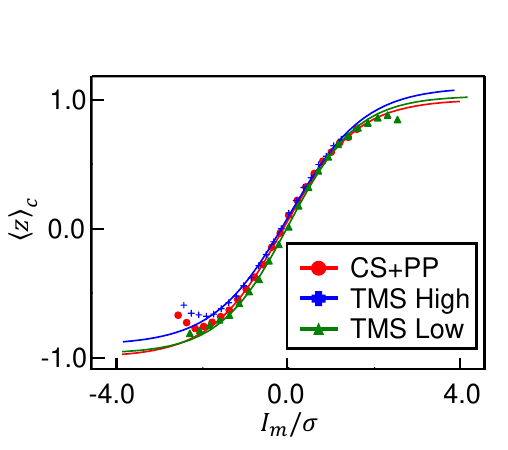}
		\caption{ \textbf{Experiment data for $z$-axis back action of weak measurement.} Tomography data for $z$-component of the qubit Bloch vector along $Q_m = -4$ of the 2D histograms shown in Fig.~\ref{fig:Science protocol}. Different color indicates the case for CS + PP, 'TMS High' match and 'TMS Low' match, respectively.}
\label{fig:Tomo_Z}
\end{figure}

For TMSL High and Low matched readout, we can use this protocol to probe the measurement back-action. While we lack a full theory for this back-action, we are encouraged by the overall outcomes resemblance to `CS+PP with altered noise'.  Indeed, in Fig.~\ref{fig:Science protocol}(b), (c) we see a close resemblance of the data to CS + PP readout. The strength of the cavity drive is varied such that the separation of the two states in the histograms is the same, so that $\bar{I}_m/\sigma$ match. This should result in a $z$ back-action which is indistinguishable for all three case.  As shown in Fig.~\ref{fig:Tomo_Z} in which we plot a line cut of $\langle z \rangle_c$ through $Q_m=-4$, this is indeed the case.  Given that the data do not vary with $Q_m$, we could also have averaged all data along $Q_m$ to achieve a similar result.  However, the $x$ and $y$ back-action, although similar in form, show very different apparent strength and efficiency, as discussed in the main text.

\subsection{Noise visibility ratio and quantum efficiency}
In this section, we will show that the observed large changes in quantum efficiency of the three readout methods \textit{can not} be explained by changes in the ratio of output noise of the TMS interferometer and the classical noise from the output chain.

\begin{figure}
	\includegraphics[scale = 1.0]{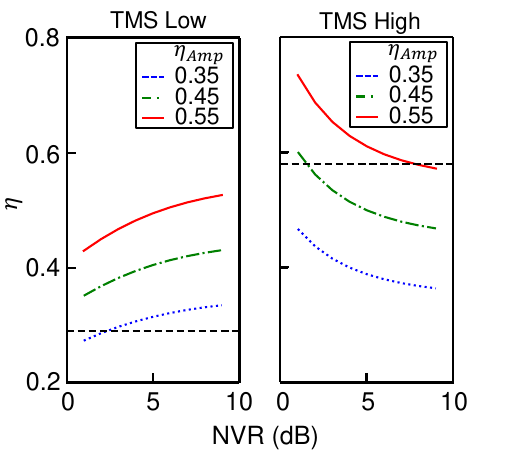}
		\caption{\textbf{Calculated overall efficiency of our measurement system for two-mode squeezed light as a function of NVR for different analyzer JPC/TMS interferometer efficiency.} Overall efficiency of the system vs NVR at the `TMS Low' and `TMS High' matched points for different amplifier efficiency $\eta_\textrm{Amp}$ = 0.35 (blue dash line), 0.45 (green dotted line) and 0.55 (red line). The black dash lines show the value of efficiency obtained from the weak measurement protocol.}
\label{fig:NVR}
\end{figure}

The overall quantum efficiency of the system extracted from the weak measurement protocol can be expressed as:
\begin{equation}
\eta = \eta_{\rm{Amp}}~\eta_{\rm{out}},
\end{equation}
where $\eta_{Amp}$ is the efficiency of the system before HEMT (qubit-cavity and JPC/TMS interferometer), $\eta_{\textrm{out}}$ is the efficiency of the output chain after the analyzer JPC which is dominated by the efficiency of the HEMT amplifier. The efficiency of the output chain, $\eta_{\textrm{out}}$, can be expressed as:
\begin{equation}
\eta_{\rm{out}} = \frac{N_{\rm{Amp}}}{N_{\rm{Amp}}+N_{\rm{out}}}.
\end{equation}
where $N_{\rm{Amp}}$ is the output noise power of the analyzer JPC, $N_{\textrm{out}}$ is the added noise power of the output chain referred back to the input of the HEMT.
An easily measurable quantity in the lab that is closely related to $\eta_{\textrm{out}}$ is the noise-visibility-ratio (NVR),
\begin{equation}
NVR = \frac{N_{\rm{Amp}}+N_{\rm{out}}}{N_{\rm{out}}}.
\end{equation}
It is easy to see:
\begin{equation}
\eta_{\rm{out}} = 1 - \frac{1}{NVR}.
\end{equation}
In our experiment, the NVR is typically 7~dB when the analyzer JPC is operated at 20~dB gain. Therefore, for coherent light readout, given $\eta = 0.46$ extracted from the weak measurement, we have $\eta_{Amp}=0.58$ for our system. 

Now consider the case of the two-mode squeezed light(TMS). Even though we did not measure the its NVR directly, we can calculate it based on the NVR of coherent light and the noise suppression/enhancement shown in Fig.~\ref{fig:Bull_eye}. The NVR for TMS at the high (H) and low (L) match points can be expressed as
\begin{equation}
{NVR}^{H/L}_{TMS} = 1 + \frac{N^{H/L}_{TMS}}{N_{\rm{out}}} = 1 + \frac{N_{CS}}{N_{\rm{out}}} (\frac{\sigma^{H/L}_{TMS}}{ \sigma_{CS}})^2,
\end{equation}
which is larger (smaller) at the high (low) match point than that of coherent light. Consequently, the efficiency of the output chain ($\eta_\textrm{out}$) will also be larger (smaller) than that of the coherent light readout.  

If we assume the quantum efficiency of TMS interferometer is the same as that in the coherent light case, namely, $\eta_{Amp} = 0.58$, and a typical coherent light NVR, 7 dB, then given the $ \frac{\sigma^{H}_{TMSL}}{\sigma_{CS}} = 1.21$ and $\frac{\sigma^{L}_{TMSL}}{ \sigma_{CS}} = 0.86$, we can calculate the overall quantum efficiency of the system at the high and low $\sigma$ match point. We get $\eta^H = 0.48 < \eta^H_{exp} = 0.58 $ and $\eta^L = 0.42 > \eta^L_{exp} = 0.29$.

More generally, in Fig.~\ref{fig:NVR} we plot the overall efficiency as a function of NVR for different values of $\eta_\textrm{Amp}$. It clearly shows that the measured efficiency at `TMS Low' and `TML High' match points requires, in addition to the changes in NVR, the analyzer JPC/TMS interferometer to have different quantum efficiencies which itself requires further studies.

\clearpage
\input{tmsl.bbl}  

\end{document}

%% file: tmsl.bbl
%